\documentclass{acmtog}

\acmVolume{VV}
\acmNumber{N}
\acmYear{YYYY}
\acmMonth{Month}
\acmArticleNum{XXX}  
\acmdoi{10.1145/XXXXXXX.YYYYYYY}

\def\runningfoot{\def\@runningfoot{}}
\def\firstfoot{\def\@firstfoot{}}

\usepackage[normalem]{ulem}

\usepackage{amssymb}
\usepackage{color}
\usepackage{listings}
\usepackage{microtype}
\usepackage{setspace}
\usepackage{multirow}
\usepackage{graphicx}
\usepackage{hyperref}

\usepackage{wrapfig}

\usepackage{tikz}

\definecolor{light-gray}{gray}{0.5}

\usepackage{algorithmic}

\input{lua.def}
\newcommand{\ic}[1]{\lstinline{#1}}

\newcommand{\name}[1]{\textbf{#1}}

\definecolor{maroon}{rgb}{0.5,0.0,0.0}
\definecolor{cardinal}{rgb}{0.77,0.12,0.23}
\newcommand{\change}[1]{#1}
\newcommand{\remove}[1]{}

\begin{document}

\markboth{G. Bernstein et al.}{Ebb: A DSL for Physical Simulation on CPUs and GPUs}

\title{Ebb: A DSL for Physical Simulation on CPUs and GPUs} 

\author{
GILBERT LOUIS BERNSTEIN
{\upshape and} CHINMAYEE SHAH
{\upshape and} CRYSTAL LEMIRE
{\upshape and} ZACHARY DEVITO
{\upshape and} MATTHEW FISHER
{\upshape and} PHILIP LEVIS
{\upshape and} PAT HANRAHAN
\affil{Stanford University}}



\acmformat{Bernstein, G. L., Shah, C., Lemire, C., DeVito, Z., Fisher, M., Levis, P., and Hanrahan, P. 2015.}

\maketitle


\begin{abstract}

Designing programming environments for physical simulation 
is challenging because simulations rely on diverse algorithms
and geometric domains.
These challenges are compounded when we try to
run efficiently on heterogeneous parallel architectures.
We present Ebb, a domain-specific language (DSL) for simulation,
that runs efficiently on both CPUs and GPUs.
Unlike previous DSLs, Ebb uses a three-layer architecture to separate
(1) simulation code,
(2) definition of data structures for geometric domains,
and (3) runtimes supporting parallel architectures.
Different geometric domains are implemented as libraries that
use a common, unified, relational data model.
By structuring the simulation framework in this way,
programmers implementing simulations can
focus on the physics and algorithms for each simulation
without worrying about their implementation on parallel computers.
Because the geometric domain libraries
are all implemented using a common runtime based on relations,
new geometric domains can be added as needed,
without specifying the details of memory management,
mapping to different parallel architectures,
or having to expand the runtime's interface.

We evaluate Ebb by comparing it to several widely used simulations,
demonstrating comparable performance to hand-written GPU code where
available, and surpassing existing CPU performance optimizations by up
to 9$\times$ when no GPU code exists.

\end{abstract}

\section{Introduction}
Physical simulation is common in computer graphics and animation, 
but building a common software infrastructure for simulation has been
difficult.  Different simulators use different numerical algorithms and 
a variety of geometric representations.
For example,
a Eulerian fluid-flow solver may use a regular grid,
a cloth simulater may use a quadrilateral mesh,
and Lagrangian finite-element code may use tetrahedral meshes.
Furthermore, simulations often involve multiple interacting physical
phenomena, requiring that the different simulators 
and geometric domains be coupled together. 
Finally, since simulation is compute intensive,
simulators must run efficiently on many different parallel architectures,
including both CPUs and GPUs.
To achieve high performance, 
simulation programmers are expected to become experts in not only
physics and numerical techniques, but also parallel programming.
As a result qualified experts are rare, making simulation code
difficult and expensive to develop.

This paper describes Ebb\footnote{
Ebb is a new implementation of the Liszt programming model
designed to support a wider variety of situations than the original
Liszt implementation.
A distribution and documentation for the language can be found at \href{http://ebblang.org}{ebblang.org}.
}, a domain-specific language (DSL)
for developing physical simulations 
of fluids and deformable meshes
that is designed to run efficiently on both CPUs and GPUs.
Ebb is motivated by the successes of prior DSLs,
such as the RenderMan shading language~\cite{renderman},
the Halide image processing language~\cite{halide},
and the Liszt~\cite{liszt} language 
for solving partial differential equations on unstructured meshes.
These DSLs use abstractions 
(lights and materials for rendering,
functional images,
and meshes/fields, respectively)
that allow simulation programmers to write code at a higher-level.
Even though DSL code is higher-level, 
the DSLs can be compiled to a wide range of computer platforms
and perform as well as code written in a low-level language.

Each of these existing DSLs 
are designed around one geometric domain
(e.g. Liszt's unstructured meshes)
whereas simulations often need to use a variety of geometric domains 
(triangle meshes, regular grids, tetrahedral volumes, etc.).
In order to support multiple geometric domains,
we propose a three-layer architecture for Ebb.
In the top layer, users write application code, 
such as a fluid simulator, 
FEM library,
or multi-physics library,
in the Ebb language using its geometric domain libraries;
Similar to shader languages,
this code says what should be computed 
for each element in a geometric domain,
capturing implicit parallelism.
In the middle layer,
different geometric domains are coded as \emph{domain libraries};
these domain libraries are implemented using a runtime API 
based on \emph{relations}.
In the bottom layer,
the Ebb runtime implements 
this relational model for each hardware target.

This three layer model has several advantages.
First, the abstraction of geometric domains into libraries provides
programmers a familiar interface to common geometric data structures.
Second, the relational abstraction is expressive enough that
different geometric domains can interoperate,
and that new geometric domain libraries can be added when needed.
Third, the relational representation allows the
the runtime to generate and execute performant parallel code.

Ebb is a first step to building an integrated simulation environement.
However, it does not yet fully support every type of simulation.
In particular, Ebb does not currently support collision detection,
nor does it support adaptive remeshing operations.
Nonetheless, Ebb does provide generic ways to interact with external libraries,
allowing users to integrate pre-existing code for unsupported capabilities,
like Fourier transforms of grid data.

In this paper, we make the following contributions:

\begin{itemize}

\item A programming model for simulation that supports a range of
  different geometric domains, plus the interactions and coupling between
  these domains.

\item A three-layer architecture that unifies multiple geometric domains
  into a common relational data model.
  This abstraction hides the complexity of 
  the underlying geometric domain library implementation
  from application programmers 
  while also ensuring that the set of supported
  geometric domains can be extended as needed.

\item A demonstration of simulations involving 
  multiple geometric domains including tetrahedral meshes,
  unstructured meshes, regular grids, and particles can be expressed,
  customized, and composed together via a common relational API.

\item An efficient implementation of the language targeting both
  CPUs and GPUs.

\item An evaluation of this programming model and its implementation
  by comparing a number of example programs for different simulations,
  including fluid simulation and deformable solids,
  to existing optimized solutions.

\end{itemize}

Our evaluation shows that Ebb performs well and is portable between
CPUs and GPUs. In all our tests
Ebb performs within 27\% of the performance of hand-tuned GPU code,
despite being more concise and easier to understand.
We also compared Ebb in cases where no GPU code was available,
finding that Ebb's GPU implementation executed 9 times faster than
a corresponding multi-core CPU implementation.  This demonstrates the
benefits of performance portability---that parallel hardware can be
exploited without additional development effort.

\section{Background}

Graphics researchers have developed a number of libraries capable of
simulating multiple physical phenomena, such as Nucleus\cite{nucleus}
and Physbam\cite{physbam}.  Ebb is not designed as a replacement
for such libraries.  Rather, it presents an alternative to directly
writing such libraries in C++.  As currently designed, multi-physics
libraries like Nucleus and Physbam are trapped by having to
support a combinatorial explosion of different code paths for simulating
different phenomena, different geometric data structures and
different parallel implementations.  In Nucleus,
\change{
(or a more recent unified particle solver
from NVIDIA~\cite{macklin2014unified})
}
this challenge is tackled
by choosing sub-optimal simulation strategies (like particle-based fluids)
in order to avoid the addition of new code paths (like Eulerian grids).
In Physbam, this challenge is tackled by the adhoc addition of code paths,
leading to a large amount of unparallelized and costly to maintain code.
  
The cost of parallelizing library-based code by hand leads to implementations with only partial support for multi-core or GPU parallelism. For instance, Vega~\cite{vega}, a FEM-based simulation library for elastic deformable solids, contains multi-threaded parallelization of some components but no GPU implementation.

Domain specific languages (DSLs) provide a way of programming at a high level, while still generating high performance parallel code for multiple architectures. In graphics, languages for rendering such as OpenGL and Renderman are widely used and researched~\cite{renderman,proudfoot,spark}. Recently there have been advances in developing DSLs for more areas of graphics. Halide and Darkroom provide a a simple, functional programming model for writing image processing pipelines~\cite{halide,darkroom}. Ebb extends this philosophy to a broader class of domains/data models for simulation in graphics.

DSLs for simulations have also been proposed in the area of scientific computing. The Liszt DSL targets partial differential equation solvers over unstructured meshes by providing high-level semantics for writing distributed computations over collections of mesh elements~\cite{liszt}. It does so by locking users into an unstructured 3D mesh domain model.
While simulations over grids and semi-structured meshes (tetrahedra), as well as 2D domains, can all technically be expressed, the resulting code
is both more difficult to write and suffers performance penalties.
This is because it is not appropriately specialized to representing the
specific kinds of topological relationships being used.

Instead of restricting ourselves to a specific data-model such as an unstructured mesh, we use \emph{relations} (the database concept) as a uniform abstraction for data models in Ebb. Previous work has shown that high-performance code can be created automatically by adapting a relational model to specific problems, such as creating optimized data structures for operating systems~\cite{hawkins2011data}.

Libraries for simulation have incorporated abstractions similar to relations. For instance,  OP2~\cite{op2} uses sets and set mappings to represent topology.
It provides a C++ and Fortran API to express simulations as parallel computations that can run on different architectures such as clusters or GPUs.  However, OP2 requires library users to hand compute relational joins between different sets. This makes it difficult to compose joins, forcing users to decompose larger computations into smaller kernels. By pre-computing all joins, OP2 also precludes optimizations for topological relations in specific domains, such as indexing arithmetic in regular grids.

Another library, Loci~\cite{loci,loci2}, uses a graph-based model to express computation over entities, where data-dependencies are expressed as rules, and functions that compute the output of a rule. Given relations and dependencies, the Loci compiler schedules computations and provides implementations that run on clusters of CPUs.  However, its declarative rules must be optimized by a query-planner, which can add overhead~\shortcite{loci2} and make performance difficult to understand. Rules are still calculated using low-level C/C++ code, so applications are currently not portable to GPUs, and the computation kernels themselves must explicitly state their dependencies. In constrast, Ebb takes a language-based approach, allowing for generation of optimized code for different architectures, and for dependencies to be inferred directly from computational kernels.

Our model of relations in Ebb extends the relational models used in OP2 and Loci by adding simulation-specific modeling primitives. By focusing these primitives on common topological access strategies used in simulations, we allow programmers to express a wide range of communication patterns while still providing clearly interpretable performance guarantees.
Furthermore, relations can be tedious to use directly and previous systems provided no way to encapsulate them into higher-level libraries. In constrast, Ebb provides a tiered approach that allows creation of re-usable domain libraries. Since these libraries are implemented using common relational operators, they can be extended or composed in the context of specific simulations.

\begin{figure}[tbp] 
  \centering
  \includegraphics{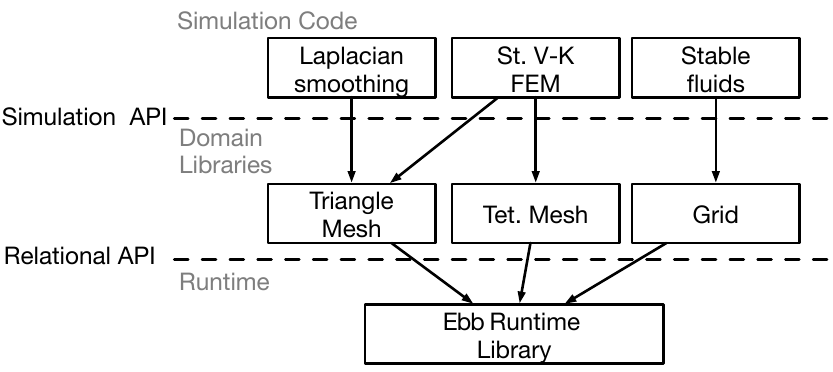}
  \caption{Ebb's three-layer architecture insulates simulation programmers
    from the definition of geometric domains by abstracting those domains
    into libraries that admit familiar syntax for the domain. The definition
    of those domain libraries are further insulated from the runtime implementation
    by encoding domains in a common, unified, relational data abstraction.}
  \label{fig:ecosystem}
\end{figure}

\section{A Three-Layer Architecture}

Usually DSLs provide a single language abstraction, separating
\emph{application code} from the language \emph{implementation}.
In previous DSLs~\cite{liszt,halide}, this abstraction is based on choosing fixed data models for computation (e.g. images, triangle-meshes,
general polyhedral-meshes).  For simulations,
which need to support a wide variety of geometric domains and
couplings between them, this approach is unwieldy, since
it requires substantial implementation effort for each domain/DSL.

Ebb resolves this problem by providing a three-layer architecture
(Figure~\ref{fig:ecosystem}) with two abstraction boundaries.
Geometric domains are abstracted into \emph{domain libraries}, whose implementation
is hidden from the simulation code that Ebb
programmers spend their time writing.  Then, these libraries are implemented
using a unified data abstraction based on \emph{relations}, separating their definition
from the runtime implementation.

At the top level, simulation programmers write code as bulk-synchronous
data-parallel kernels launched over sets of elements in the domain,
similar to vertex or pixel shaders.  Within these kernels, neighboring
elements and field data are accessed using familiar syntax such as \ic{e.head}
for the vertex at the head of an edge and \ic{e.head.pos} for the position
of that vertex.  This allows the programmer to write local
computations ranging in complexity from a step of Jacobi iteration
to the assembly of a stiffness matrix for FEM simulation.  By sequencing
these kernels, a full simulation is executed.

At the middle level, developers of domain libraries implement or customize
common geometric domains like triangle and tetrahedral meshes, 2D/3D
regular grids, or general unstructured polyhedral/polygonal meshes.  These
libraries (about 100-400 lines of code each) encode the topology of the
geometric domain using a relational data API.  This abstraction
allows for the set of domain libraries to be extended, (without
modifying the runtime) but still ensures that critical information about
the connectivity and structure of the data is provided to the runtime.

At the bottom level, the implementers of Ebb provide the language and
relational API implementations, including backends for targeting both
CPU and GPU.  By exploiting code analysis and the abstract specification
of data, the runtime is able to manage data movement, adjust data-layouts
to different architectures, and re-implement primitives like reductions
using machine-appropriate techniques.

The next three sections look at each of these abstraction levels.

\section{Simulation Programming (by example)}
\label{sec:programming}

Ebb programs are organized around the observation that simulation
execution time is dominated by data-parallel operations,
which we call \name{kernels}.
Consequently, we can logically decompose Ebb programs
into top-level control code that orchestrates the simulation and
the data-parallel kernels which perform the bulk of the work.
Ebb reflects this dichotomy by embedding an Ebb kernel language in
Lua~\cite{eyeofneedle}, a scripting language popular in the
game development community.  Ebb code handles data parallel computation,
while the Lua code orchestrates the control.
In the following, we illustrate the
design of this language through example simulation code.

\begin{figure}[t]

\begin{tikzpicture}[overlay, x=3.5pt, y=7.25pt, shift={(1.8,-1)}]
  \path[rounded corners=3.5pt, fill=magenta!15!white]
    (-0.5, 0.2)  rectangle (66.5,-2.2);
  \path[rounded corners=3.5pt, fill=cyan!15!white]
    (-0.5,-2.8)  rectangle (66.5,-13.2);
  \path[rounded corners=3.5pt, fill=yellow!15!white]
    (-0.5,-13.8) rectangle (66.5,-37.2);
  \path[rounded corners=3.5pt, fill=lime!15!white]
    (-0.5,-37.8) rectangle (66.5,-50.2);

  \draw[magenta, thick, rounded corners=3.5pt]
    (-0.5, 0.2)  rectangle (66.5,-2.2);
  \draw[cyan,    thick, rounded corners=3.5pt]
    (-0.5,-2.8)  rectangle (66.5,-13.2);
  \draw[yellow,  thick, rounded corners=3.5pt]
    (-0.5,-13.8) rectangle (66.5,-37.2);
  \draw[lime,    thick, rounded corners=3.5pt]
    (-0.5,-37.8) rectangle (66.5,-50.2);

  \node [below left] at (66,  0) {(1)};
  \node [below left] at (66, -3) {(2)};
  \node [below left] at (66,-14) {(3)};
  \node [below left] at (66,-38) {(4)};
\end{tikzpicture}

\begin{lstlisting}
local Tetmesh = L.require 'domains.tetmesh'
local dragon  = Tetmesh.Load('dragon.off')

local K  = L.NewConstant(L.float, 1.0)
local dt = L.NewConstant(L.float, 0.0001)
local E  = L.NewGlobal(L.float, 0.0)
dragon.edges:NewField('rest_len', L.float):Load(0)
dragon.vertices:NewField('mass', L.float)
               :Load('dragon_mass.data')
dragon.vertices:NewField('q', L.vec3f)
               :Load(dragon.vertices.pos)
dragon.vertices:NewField('qd', L.vec3f):Load(...)
dragon.vertices:NewField('force', L.vec3f):Load({0,0,0})

ebb kernel initLen( e : dragon.edges )
  var diff   = e.head.pos - e.tail.pos
  e.rest_len = L.len(diff)
end

ebb kernel computeInternalForces( v : dragon.vertices )
  for e in v.edges do
    var dq    = e.head.q - v.q
    var dir   = L.normalize(dq)
    v.force  += K * (e.rest_len * dir - dq)
  end
end

ebb kernel applyForces( v : dragon.vertices )
  var qdd = v.force / v.mass
  v.q    += v.qd * dt + 0.5 * qdd * dt * dt
  v.qd   += qdd * dt
  v.force = {0,0,0}
end

ebb kernel measureTotalEnergy( v : dragon.vertices )
  E += 0.5 * v.mass * L.dot(v.qd, v.qd)
end

initLen(dragon.edges)

for i=0, 10000 do
  computeInternalForces(dragon.vertices)
  applyForces(dragon.vertices)

  if i % 1000 == 999 then
    E:set(0)
    measureTotalEnergy(dragon.vertices)
    print('energy: ', E:get())
  end
end
\end{lstlisting}
\caption{A Spring-Mass simulation written in Ebb.}
\label{example1}
\end{figure}

\subsection{A Spring-Mass Simulation}
\label{sec:spring-mass}

To begin this program, we load in a tetrahedral mesh of a dragon.
(Figure~\ref{example1}.1) Like all tetrahedral meshes, this dragon
comes equipped with unordered sets of elements
(\ic{vertices}, \ic{edges}, \ic{tetrahedra}, etc.)
that we refer to as \name{relations}, ie. relational tables.

Once the basic domain is loaded, we can extend it
(Figure~\ref{example1}.2) with additional \name{fields} of data,
\ic{rest_len}, \ic{mass}, \ic{q}, \ic{qd}, \ic{force}.  These fields
are defined over relations, specifying a value of the field for each
element of the relation.  For instance, the last declaration says
``There is a vector of 3 floats for each vertex, called `force';
initialize it to the zero vector.''  In addition to these fields, the
programmer can further define constant and \name{global} values that do
not vary across the domain.  GPU programmers will recognize these as
similar to uniform data in shader languages.

Since all of this setup is part of the control logic, it is written in
Lua (here the \ic{local} keyword defines new variables).
In the next segment of the code, (Figure~\ref{example1}.3)
we define the kernels using the Ebb kernel language, as indicated
by the prefix keywords \ic{ebb kernel} and use of the keyword \ic{var}
to declare variables.

The spring-mass simulation defines four Ebb kernels 
(Figure~\ref{example1}.3), which are data-parallel computations that can be executed
for each element of the relation indicated in the type annotation
(e.g. \ic{e : dragon.edges}).
Within each kernel, we can access the fields we defined as members of
elements, e.g. \ic{e.rest_len}, reading from, writing to,
and reducing into them\footnote{Ebb supports the reduction operators
\ic{+=}, \ic{*=}, \ic{max=}, and \ic{min=}}.
Additionally, the tetrahedral mesh comes equipped with pre-existing
topological connections, which we can use to access neighboring elements,
and fields defined on those elements.  For instance, the kernel
\ic{initLength} looks up the initial position of the head and tail
vertices of an edge (\ic{e.head.pos}, and \ic{e.tail.pos}) in order
to compute the resting length of the spring corresponding to that edge.

In \ic{computeInternalForces()} we see another idiom for neighbor access,
where we loop over the set of (directed) edges leaving a vertex.
Using this pattern, we can compute the total force exerted on a
vertex by the (variable number of) edge-springs it's connected to.

The Ebb model prevents race conditions during a parallel kernel by enforcing that a field is in a single \name{phase} during each kernel. A field can be in a \emph{read-only} phase, a \emph{reduction} phase with a particular reduction such as addition, or an \emph{exclusive access} phase where the fields of the parameter element (e.g. \ic{v} in \ic{applyForces()}) can be read and written exclusively by the instance of the kernel handling that element. This safety mechanism is reflected in this example where we have one kernel \ic{computeInternalForces()} to update the spring forces in \ic{v.force} (using  a reduction) and another kernel \ic{applyForces()} to read these forces and reset them (using exclusive access). If we tried to merge these two into a single kernel, Ebb would report a phasing error since \ic{v.force} would be used in two different phases.

Ebb ensures \emph{performance portability}
by restricting the ways that programmers can access memory from within
kernels and by preventing race conditions.  This guarantees that
code developed and debugged while running single-threaded on a CPU can
safely be ported to GPUs and other parallel architectures.
While the kernel language is a standard imperative
language, the programmer must access all data via local connections
from the parameter element.  This restriction is what gives the runtime
sufficient information to reason about how to best map the computation
to different parallel architectures.

Finally, our program launches the kernels (Figure~\ref{example1}.4)
from a simulation loop written in Lua.  This results in a familiar,
bulk synchronous parallel execution model.

\begin{figure}[t]

\begin{tikzpicture}[overlay, x=3.5pt, y=7.25pt, shift={(1.8,-1)}]
  \path[rounded corners=3.5pt, fill=cyan!15!white   ]
    (-0.5,-2.8)  rectangle (66.5,-9.2);
  \path[rounded corners=3.5pt, fill=yellow!15!white ]
    (-0.5,-9.8) rectangle (66.5,-21.2);
  \path[rounded corners=3.5pt, fill=lime!15!white   ]
    (-0.5,-26.8) rectangle (66.5,-29.2);

  \draw[cyan,    thick, rounded corners=3.5pt]
    (-0.5,-2.8)  rectangle (66.5,-9.2);
  \draw[yellow,  thick, rounded corners=3.5pt]
    (-0.5,-9.8) rectangle (66.5,-21.2);
  \draw[lime,    thick, rounded corners=3.5pt]
    (-0.5,-26.8) rectangle (66.5,-29.2);

  \node [below left] at (66, -3) {(1)};
  \node [below left] at (66,-10) {(2)};
  \node [below left] at (66,-27) {(3)};
\end{tikzpicture}

\begin{lstlisting}
-- Load Grid; Define Diffusion, Projection, Advection
...

local particles = L.NewRelation {
  name = 'particles',         size = M,
}
particles:NewField('dual_cell', grid.dual_cells):Load(...)
particles:NewField('pos', L.vec3f):Load(...)
particles:NewField('vel', L.vec3f):Load(...)

ebb kernel update_particle_vel( p : particles )
  var x1 = fmod(p.pos[0] - 0.5f)
  var y1 = fmod(p.pos[1] - 0.5f)
  var x0 = 1.0f - x1
  var y0 = 1.0f - y1

  p.vel = x0 * y0 * p.dual_cell.cell(0,0).vel
        + x1 * y0 * p.dual_cell.cell(1,0).vel
        + x0 * y1 * p.dual_cell.cell(0,1).vel
        + x1 * y1 * p.dual_cell.cell(1,1).vel
end

for i=0, 10000 do
  ...
  update_particle_vel(particles)
  update_particle_pos(particles)

  grid.dual_cells:PointLocate(particles.dual_cell,
                              particles.pos)

end
\end{lstlisting}
\caption{Coupling particles to a fluid simulation.}
\label{example3}
\end{figure}

\subsection{Coupling Domains}

Many simulations use more than one geometric domain, so Ebb also includes support for coupling domains together. As a simple example of coupling domains, we show how tracer particles
can be added to a simple \change{Stable Fluids\cite{stam1999stable}}
simulation (Figure~\ref{example3}).
For brevity, we omit the majority of the simulation code and focus solely
on the code demonstrating the interaction between the grid and particle
domains.  To couple two domains, we need a way to establish
a coupling, use that coupling, and update the coupling.

\begin{wrapfigure}[8]{l}{0.75in} 
  \vspace{-3mm}
  \includegraphics[width=1.0in]{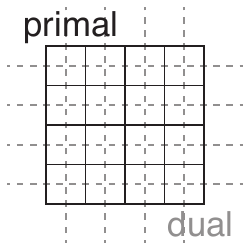}
  \vspace{-4mm}
\end{wrapfigure}
In order to \emph{establish} a coupling, we define a field typed by the
relation we want to connect to.  For instance, here
(Figure~\ref{example3}.1) we give each particle a reference to the
dual-cell of the grid in which it's currently located.

Given this link, we can \emph{use} it to linearly interpolate
(Figure~\ref{example3}.2) the velocity field stored at the
cell-centers of the mesh, deriving velocities with which to advect
the particles.  Since these cells are in a regular grid domain, we can access them via relative indexing operators, e.g. \ic{cell(0,1)}.

After having updated the particles' positions, we want to similarly
\emph{update} the coupling to reflect the particles' new positions.
We do so using the \ic{PointLocate()} function (Figure~\ref{example3}.3).
For each particle, the appropriate dual cell is computed using the
supplied position field.
While the current implementation of Ebb is designed for
simulations with static topology, 
we found that adding point location in a grid was trivial in the current design, and enabled a range of useful techniques in fluid simulations. 
Besides locating particles in a grid, this same mechanism can be used
to perform the semi-Lagrangian advection step of Stable Fluids simulations.

\subsection{Code Interoperability}

Similar to other graphics languages, Ebb is developed with the intention that it be used as a part of a larger application. To this end, Ebb is written as an embeddable Lua library that can be linked into existing applications. It is also able to provide low-level views of its simulation memory so that other libraries do not need to marshal data out of Ebb, enabling easy interoperability with existing code and avoiding the costs of memory duplication and copying.

Ebb also allows the simulation programmer to call out to external library functions
to operate on Ebb data in-between kernel invocations. This is valuable when at each
iteration the simulation needs to invoke transforms that do not map well to Ebb's
computational model, such as collision detection or the discrete Fourier transform.
For example, a simulation programmer could invoke external libraries such as CUFFT\footnote{developer.nvidia.com/cuFFT} or FFTW\footnote{fftw.org/} without the need to marshal the data or escape the simulation. This approach to interoperability follows many of the design patterns used in other
embedded languages such as Lua and OpenGL.

\section{Developing Domain Libraries}
\label{sec:domain_modeling}
\begin{figure}[htbp] 
  \centering
  \includegraphics[width=3.25in]{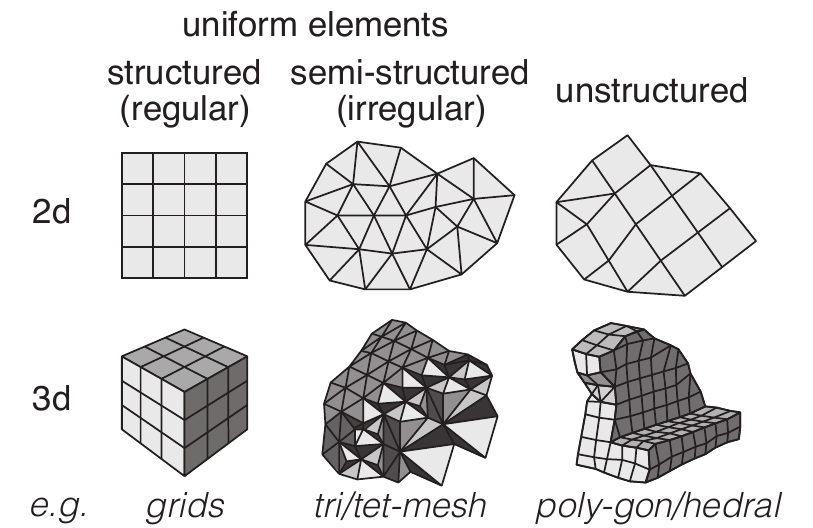}
  \caption{ \change{
    Ebb's relational model is sufficiently expressive to encode (at least) all of the geometric domains in this taxonomy, while also taking advantage of specific details like dimensionality or regularity. (Ebb can also efficiently express other domains, such as particles.)
  } }
  \label{fig:taxonomy}
\end{figure}

In this section we describe how geometric domains such as the tetrahedral mesh used in the spring mass example are implemented as libraries.  That is, we describe the set of primitive operations that we use to build the existing domain libraries.  By definining and describing these primitives, we make it possible for new domain libraries to be written, resulting in an extensible set of geometric domains for simulations.
These data modeling primitives both need to be expressive enough to
construct and couple a range of different domains,
(at least those described in the taxonomy of figure~\ref{fig:taxonomy})
and also informative
enough that we can efficiently implement the model on a range of different
architectures.

To better understand why data abstraction is necessary, consider two
strawman proposals: express topological relationships by storing data
in arrays and identifying elements with array indices, or storing structs
of data connected via pointers in the heap.  Both of these proposals are
too low-level.  They lose critical typing information about what sort of
data is accessed, and overcommit the runtime to particular data layouts.
Data referenced by a pointer cannot be moved, and indexed arrays cannot
be re-ordered without potentially breaking client code.

Instead, Ebb relies on the relational concept of a \name{key}, similar to
references in garbage collected languages.  Keys provide an opaque
reference to an element of a particular relation.  Kernels are not
allowed to perform arithmetic on keys, nor otherwise rely on their
encoding.  As a result, the runtime is
(1) able to jointly analyze the kernel code together with the loaded data,
and (2) reserves the ability to optimize data layout.
As a stark example of the benefits, Ebb is able to guarantee
that most of its memory accesses will not segfault by construction,
whereas general aliasing analysis on pointers or numeric indices is
undecidable.

Furthermore, we observe that the topological relationships 
defining geometric domains for simulation
can be boiled down to a few cases: one-to-one relationships like
defining the head corresponding to each edge, one-to-many relationships,
like getting all the edges or triangles touching a vertex, and arithmetical
offsets, like accessing a stencil pattern around a grid cell.

Ebb encodes these relationships using three primitives: key-fields,
query-loops, and affine-indices.  Respectively, these allow the encoding
of one-to-one functions between relations, inversions of those functions,
and offsets within a grid.  By further using the common database pattern
of auxiliary tables, we can build one-to-many relationships out of
one-to-one functions and their inverses.

By relying on a restricted set of primitives,
Ebb provides a relational data model with substantially
more predictable performance than SQL or relational algebra.
Unlike with general database modeling primitives, our more
\change{specialized primitives} guarantee that all topological
accesses will have constant-cost, or even zero-cost in terms
of required memory accesses.

\subsection{Relational Modeling Primitives (by example)}

As a running example, we will explain the construction of a standard
triangle mesh domain.  To begin, we create three relational tables
to model the triangles, (directed) edges, and vertices of the mesh.

\vspace{0.1in}
\centerline{
\includegraphics[width=2.5in]{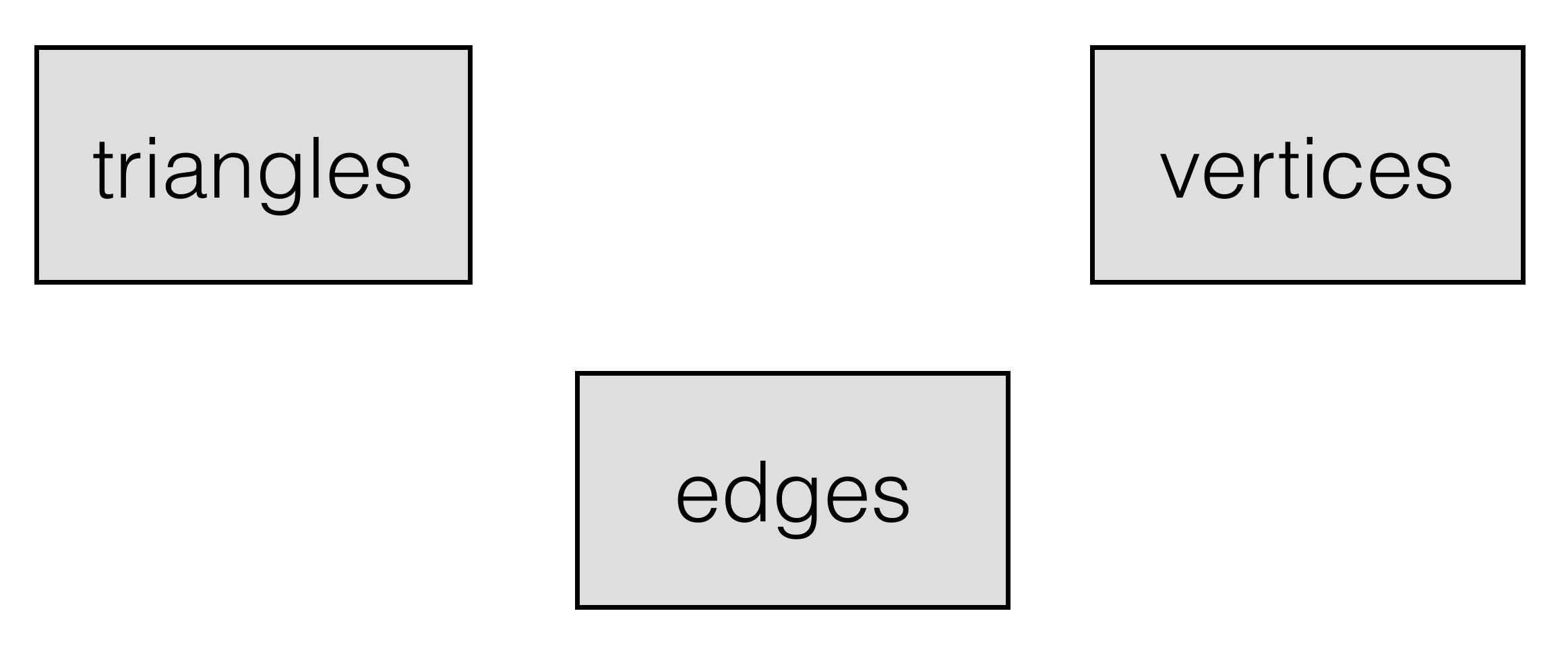}
}

\begin{lstlisting}
triangles = L.NewRelation { name='triangles', size=n_tri  }
edges     = L.NewRelation { name='edges',     size=n_edge }
vertices  = L.NewRelation { name='vertices',  size=n_vert }
\end{lstlisting}

\name{Key-fields} encode functional relationships like the \ic{head} or
\ic{tail} of an edge; that is, each edge defines exactly one vertex that
is the head or tail, respectively.  Key-fields are set up by declaring a
new field typed by some relation, and loading in a list of initial
values\footnote{ While the numeric value of these keys is important
for initializing the relationship, Ebb is subsequently free to reorder
the relations, encode keys via pointers, or use any other choice of
implementation strategy. }.
Earlier code examples have provided many examples of key-fields being
used by simulation code.

\vspace{0.1in}
\centerline{
\includegraphics[width=2.5in]{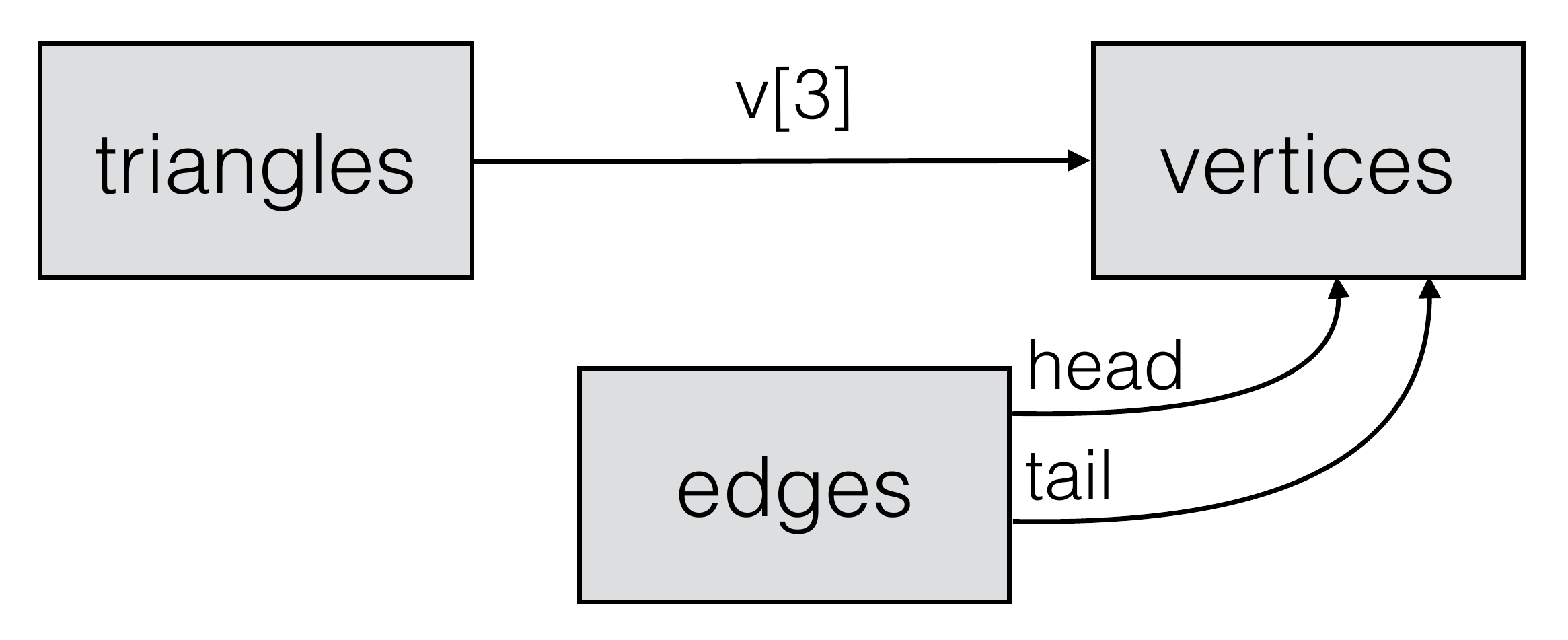}
}

\begin{lstlisting}
triangles:NewField('v', L.vector(vertices, 3)):Load(...)
edges:NewField('head', vertices):Load(...)
edges:NewField('tail', vertices):Load(...)
\end{lstlisting}

\name{Query-loops} enable one-to-many relationships to be expressed.
They encode access to the inverse of a key-field relationship.
For instance, the query-loop expression \ic{L.Where(edges.tail, v)}
gets a list of edges \ic{e} where \ic{e.tail == v}.
For this expression to be valid, we require that it is set up
beforehand using an API call to \name{group} the target relation
(here \ic{edges}) by the field we want to invert (\ic{tail}).
This grouping allows our runtime to do preprocessing to
make the query loops efficient during execution.  

To hide the details of query loops from the simulation API,
we include the ability to define simple macros on elements.
These macros allow simulation programmers to write an intuitive
loop expression (\ic{for e in v.edges do ... end}) since
\ic{v.edges} is defined as a macro that expands into
the \ic{L.Where} query expression:

\vspace{0.1in}
\centerline{
\includegraphics[width=2.5in]{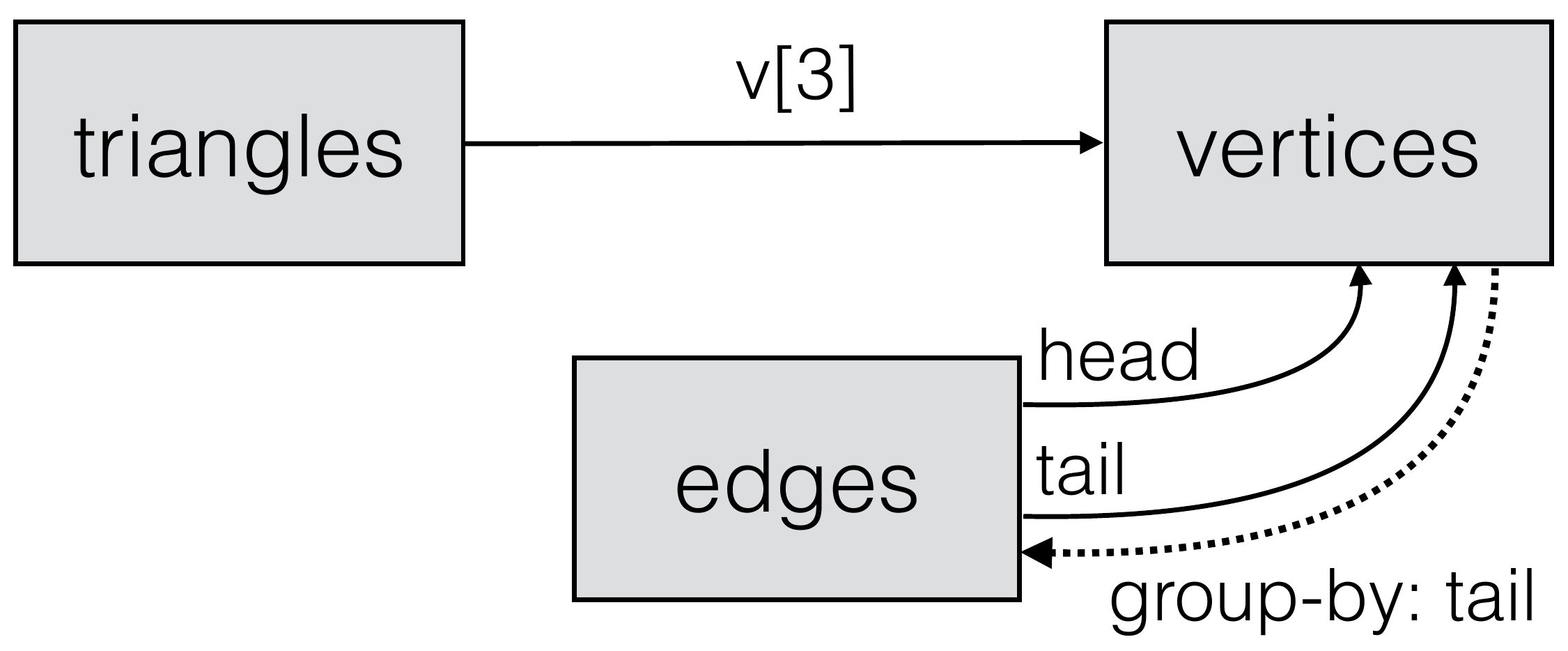}
}

\begin{lstlisting}
edges:GroupBy('tail')
vertices:NewFieldMacro('edges', L.NewMacro(function(v)
  return ebb `L.Where(edges.tail, v)
end))
\end{lstlisting}

We can compose the two features of key-fields and
query-loops, in conjunction with auxiliary tables to express arbitrary
one-to-many relationships, like all of the triangles touching a given
vertex.  We simply create a table containing all touching
triangle-vertex pairs, (i.e. a relation with 2 key fields)
and group it by the vertices.
We can even view the edges as a kind of instance of this pattern, encoding
an arbitrary many-to-many relationship from vertices to themselves.
This leads to the observation that any graph, directed, undirected,
with or without repeated edges and/or self-loops can be encoded using
only key-fields and query-loops.

\subsubsection*{Affine-indexing for regular grids} 

In our implementation of the regular grid domain, we provide multiple kinds of elements in our grid (e.g. vertices, cells), allowing users to store quantities on each:

\begin{lstlisting}
cells    = L.NewRelation { name='cells',    dims={nX,nY} }
vertices = L.NewRelation { name='vertices', dims={nX+1,nY+1} }
\end{lstlisting}

To link these elements together, we also provide an explicit
\name{affine-indexing} operator to encode topological relationships
between elements in regular grids that can be computed arithmetically
rather than looked up in memory. To capture constant relative offsets,
and simple down/up-sampling patterns, we allow the library author
\begin{wrapfigure}[9]{r}{0.75in} 
  \vspace{-5mm}
  \hspace*{-10mm}
  \includegraphics[width=1.25in]{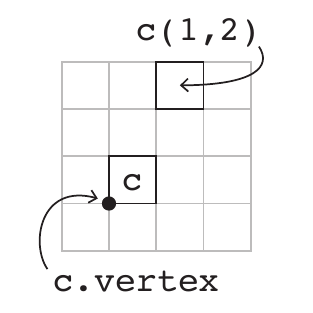}
\end{wrapfigure}
to request any affine transformation of grid indices that
they wish\footnote{Affine transforms were chosen as the most
general transformation which is closed under composition and
amenable to being bounded by compilers' shape analyses.}.
Again, we use macros to hide the verbose, but
general purpose \ic{L.Affine} primitive from the simulation programmer.
\change{Instead we expose convenient syntax like \ic{cell(1, 2)}
(the \ic{__apply_macro} macro) for accessing the neighboring cell that is
one cell to the right and two cells up, or \ic{c.vertex}
for retrieving the vertex to the bottom left of a given cell.}

\begin{lstlisting}
cells:NewFieldMacro('__apply_macro', L.NewMacro(function(c,x,y)
  return `L.Affine(cells, {{1,0,x},
                           {0,1,y}}, c)
end))
cells:NewFieldMacro('vertex', L.NewMacro(function(c)
  return `L.Affine(vertices, {{1,0,0}
                              {0,1,0}}, c)
end))
\end{lstlisting}

Ebb also includes a few other mechanisms like \name{subsets}
(e.g. interior and boundary cells) and \name{periodicity} that can be used
to encode common boundary conditions.  They work as you might expect, but
are mostly orthogonal to the core design issues we deal with here.

\subsection{Customized Domain Modeling}

While Ebb provides a standard library of domains suitable for most
common simulations, the relational abstraction primitives make the
customization, coupling and/or construction of domains relatively easy.
The following examples can be written in 100 lines of code or less,
most of which is devoted to data marshaling in Lua.

\paragraph*{Example: FEM tetmesh}
One of the applications we evaluate in our results (\S\ref{sec:results}) is an FEM simulation on a tetrahedral mesh domain.  Here we explain the modeling of this domain in detail in order to illustrate the flexibility of the relational primitives and the value of being able to customize a domain to match a simulation.

\begin{figure}[htbp] 
  \centering
  \includegraphics[width=2.5in]{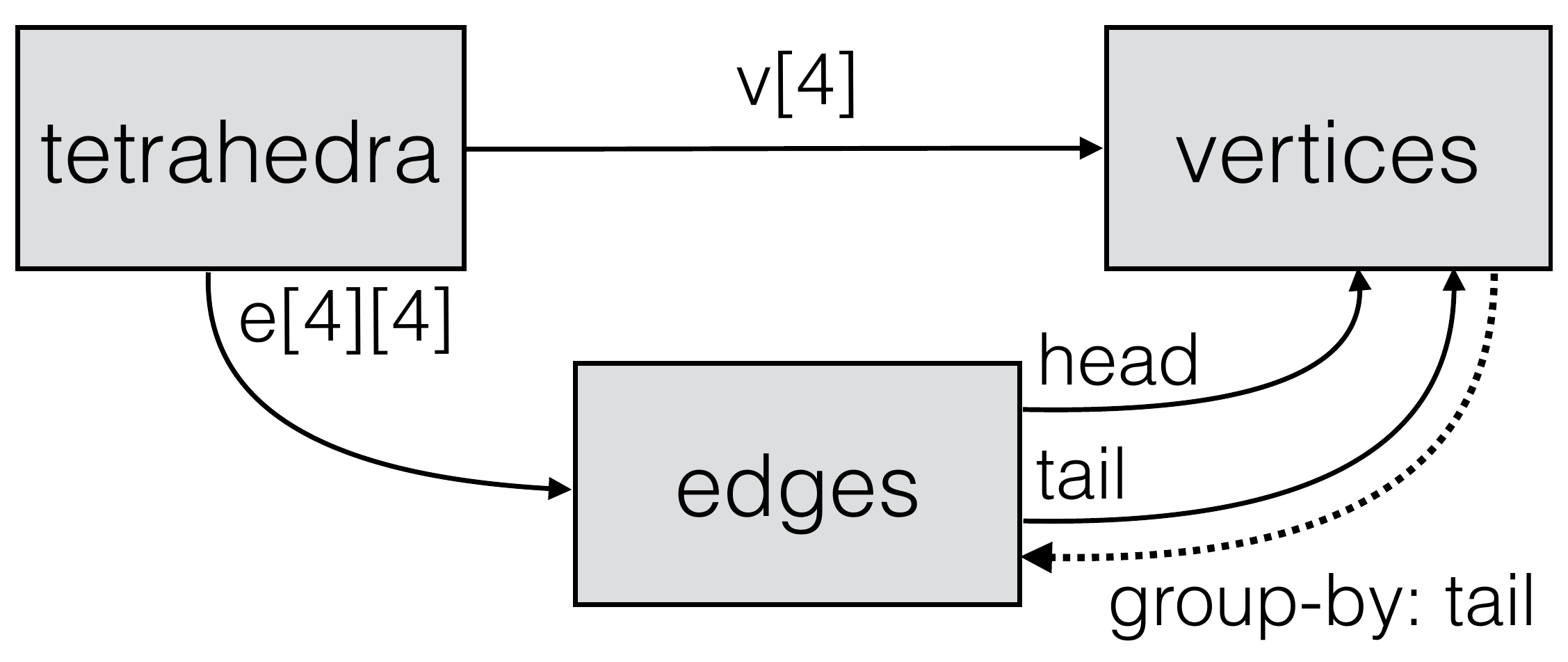}
  \caption{The tetrahedral mesh domain we use for our FEM example is unlikely to be provided in standard libraries.  The edges relation includes not only directed edges, but also a self-loop edge for each vertex, in order to model the support structure of the sparse stiffness matrix (rather than the geometric concept of an edge).  By further addressing the edges of a tet using a $4\times4$ matrix of keys, we make it easy for simulation code to update the stiffness matrix.}
  \label{fig:tetmesh}
\end{figure}

Similar to the triangle mesh,
the tetrahedral mesh has 3 relations, \ic{tets}, \ic{verts}, and \ic{edges}.  
Key-fields are defined for tet vertices \ic{tets.v[4]}, and edge endpoints \ic{tail} and \ic{head}.  We group the edges by \ic{tail}.
We make the choice to include not only directed edges, but also a self-loop edge per vertex.  Rather than being principally geometric, edges of the tet mesh are used to model the support structure for sparse matrices on the domain, and these self-loops correspond to diagonal entries.
Since updates to the edge operator are computed per-tetrahedron and reduced into the appropriate edges, we further augment the tet relation with a matrix-organized key-field \ic{tets.e[4][4]} simplifying indexing code.  To encode the stiffness matrix itself, we store $3\times3$ matrix values per-edge in a stiffness field, producing, in aggregate, a fairly sophisticated sparse block matrix encoding that can be easily addressed and updated from the tetrahedra.

\paragraph*{Example: rendermesh coupling}
Since geometric domains in Ebb are built out of a collection of relational tables, there is no reason that simulation code can't link relations of different domains together in a given simulation.  For instance, a common strategy for soft-body FEM simulations in graphics is to embed a high resolution triangle mesh of an object inside of a lower resolution tetrahedral mesh, on which the simulation actually occurs.  To do this in Ebb, a programmer only needs to set up one additional key-field \ic{trimesh.vertices.tet} of \ic{tetmesh.tetrahedra}-type.  Then, vertices of the triangle mesh can update their position by interpolating the positions of the containing tet's vertices \ic{v.tet.v[k].pos} for \ic{k=0,1,2,3}.

\begin{figure}[htbp] 
  \centering
  \includegraphics{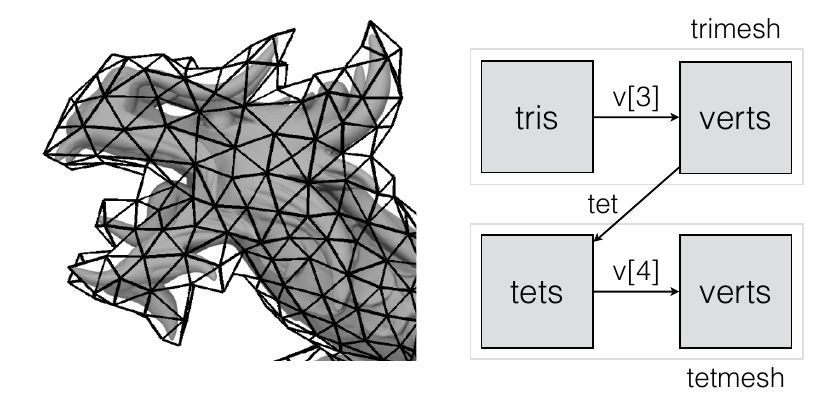}
  \caption{Because all domain data is relational, programmers can easily wire together multiple domains when setting up a simulation.  For instance, a triangle mesh intended for rendering can be embedded in a tetrahedral simulation mesh.  The relational format of the triangle mesh can then be fed directly into the rendering pipeline.}
  \label{coupling}
\end{figure}

\section{Runtime Implementation}
\label{sec:implementation}

Like OpenGL, the Ebb runtime is implemented as a series of API functions to do things such as register new relations, add fields, or register new kernels. We use Lua to expose these API functions. Ebb kernels are embedded in Lua code and are registered with the Ebb API. This is similar to how OpenGL shader programs can be embedded (as strings) in a C program and then loaded into the OpenGL API. The difference is that we use a syntax extension to embed Ebb kernels rather than using strings.

\begin{figure}[th]
\includegraphics{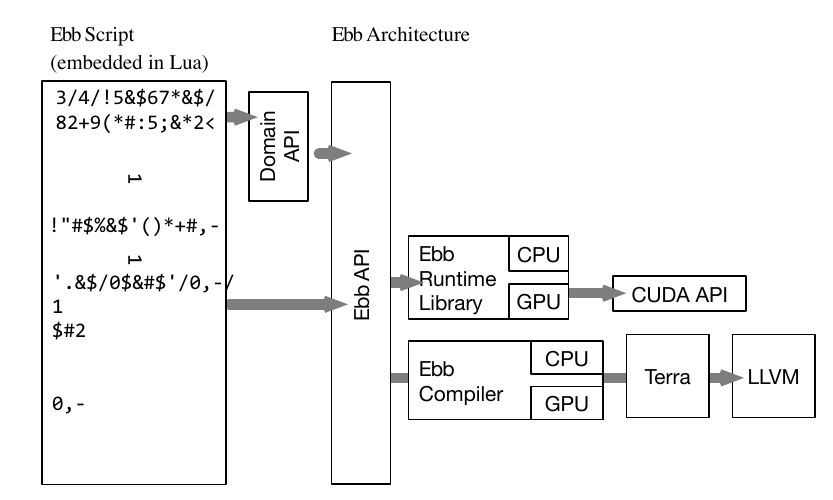}
\caption{Ebb architecture: Ebb is a language embedded in Lua. API calls are immediately executed by the Ebb runtime. Ebb kernels are compiled to CPU or GPU using Terra.}
\label{architecture}
\end{figure}

Figure~\ref{architecture} summarizes this API design. The three-layer design of Ebb allows us to keep the actual runtime API relatively simple since domains like triangle meshes or regular grids are expressed in terms of relational operators.  The runtime of Ebb only needs to implement support for this lower relational level. It needs to provide: (1) a \emph{runtime library} that includes the implementation of the relational primitives such as creating new relations, fields, and grouping them, and (2) a \emph{compiler} that translates Ebb kernels into machine code that runs on CPUs or GPUs, which includes the implementation of language features such as reductions.

\change{
In the interest of brevity, we focus the rest of this section on those implementation details that differ from standard compiler writing practice.
}

\subsection{Runtime library for relational operators}
 
\paragraph*{Data Layout and Management}
Ebb uses a column-store layout for its tables (commonly referred to as struct-of-arrays).  That is, each field is allocated a separate contiguous array of memory.
Loading data from columnar fields is more likely to result in coallesced memory loads on the GPU, boosting performance significantly on regular domains.
Field data is transfered from CPU to GPU automatically depending upon the needs of the kernel being run. User code that puts data into and gets data out of the API does not have to change to accommodate different data locations.

\paragraph*{Relational Primitives Implementations}
The data stored in fields is also used to implement the modeling primitives defined in the previous section. \emph{Key-fields} are stored as normal fields with underlying C-type \ic{uint64}.  \emph{Affine indexing} is compiled to arithmetic expressions of ID of each relational element and thus does not require any data storage. For multi-dimensional keys, modulus operations are used to extract a row's \ic{x}, \ic{y}, and \ic{z} components.  Likewise, \emph{point location} in a structured grid can be implemented via arithmetic on the coordinates and the point.

Support for one-to-many relationships are provided via \emph{query-loops}, which require the use of the \ic{relation:GroupBy(...)} API call. \change{Recall that we group a relation by one of its (key-)fields (e.g. grouping the \ic{edges} by their \ic{tail}). We call the grouped relation the \emph{target} and the relation that the key-field refers to the \emph{source}  (e.g. \ic{edges} is the target and \ic{vertices} the source).  Then in order to implement the grouping operation, we first sort the target relation by the field of keys; and second we add a hidden \emph{index} field to the source relation, storing the range of rows in the target relation whose key-field value matches that row. (The range can be encoded as two indices.)  In effect, this gives us an inverted index for the original key-field relationship in which we can directly look up the loop iteration bounds when executing a query-loop for a given row.}


\begin{figure}[thbp] 
  \centering
  \includegraphics[width=3.25in]{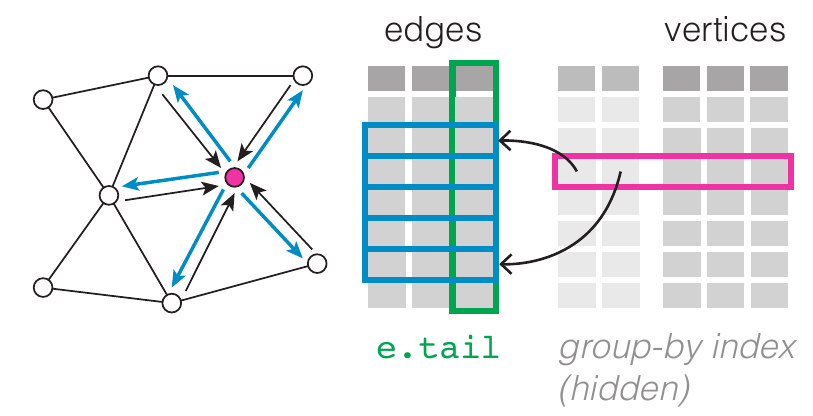}
  \caption{\change{
  When the (directed) edges of a triangle mesh are grouped by their \ic{tail} field, which contains vertex keys, the runtime (1) sorts the edges relation/table by the \ic{tail} field/column and (2) constructs a hidden index parallel to the vertices relation/table.  This index allows us to efficiently execute query-loops by  simply looking up the loop iteration bounds for each vertex.
  }}
  \label{trimesh}
\end{figure}

This approach has the same overhead as storing explicit adjacency lists (e.g. for each vertex storing a list of triangles it is adjacent to), and is equivalent to the compressed row storage layout for sparse matrices, which is known to be efficient. By abstracting this best practice, our relational design also ensures that this data is always valid and used correctly in each loop.

\subsection{Compilation of Ebb Kernels}

The runtime also includes a compiler that translates high-level Ebb kernels into machine code. Like OpenGL, this process is done dynamically in the running process so that the user does not need to run a separate compilation step beforehand. We use the Terra~\cite{terra} language framework to do this dynamic compilation. Terra is a low-level language embedded in Lua that makes it easy to create DSLs such as Ebb.

Once handed to the Ebb API, kernels are type-checked, compiled, and then executed. In addition to standard type-checking, Ebb also checks that the fields in each kernel are only used in one particular phase (\emph{read-only}, \emph{reduction} by a particular operator, or \emph{exclusive access}), using a process similar to DeVito et al.~\shortcite{liszt}. 

The code is then compiled for a particular architecture. To compile the kernel, we use the multi-stage programming facilities in Terra to generate low-level C-equivalent code implementing the kernel.  Depending on whether data is CPU or GPU resident, this kernel code is specialized for CPU or GPU execution (details below). Terra itself uses LLVM as its backend, targeting GPUs using NVIDIA's NVPTX (the device-independent CUDA target language) and CPUs using the x86 code generation paths.

\paragraph*{CPU/GPU Specialization}
Ebb maintains separately specialized, compiled functions for execution on CPU and GPU.  On CPU, the kernel body code is wrapped in a for loop, which iterates over the row indices of the argument relation.  Key variables are compiled into these index values, (\ic{uint64}) which are then used to index into various field arrays.  On the GPU, the process is nearly identical, except the for loop is replaced with a CUDA kernel launch.  CUDA kernels are launched with 64 threads per block by default, using a linear key space.

\paragraph*{Reductions}
Reads and writes can be trivially translated into load and store statements, but special care is necessary to handle reduction statements in Ebb when code is run in parallel on the GPU. Reductions fall into two categories: (1) reductions to an element of a \emph{field}, and (2) reduction to single \emph{global} value.  In the first case of field reductions, relatively few threads will write to one entry, so we use intrinsic GPU atomic reductions to implement them.  In the few cases where the GPU does not have an equivalent atomic reduction, then we use the compare-and-swap instruction to emulate it.

The second case of \emph{global} reductions is implemented using a variant of the two-pass reduction described in \cite{cudahandbook}. The algorithm is modified in two ways so that the first-pass kernel can be fused into the original Ebb kernel code:  (1) Rather than read the data to be reduced from global memory, reduction operations inside the kernel body directly reduce values into a per-block region of shared-memory.  (2) Code is appended to the end of the kernel to reduce this shared memory array and write the result to global memory.  The second kernel is then immediately launched to reduce the per-block values into the global value using the reference algorithm identically.

\section{Performance Evaluation}
\label{sec:results}

Ebb aims to capture a wide range of simulation domains, produce high performance code, and support interoperation with existing libraries.
To evaluate whether we achieved these goals, we selected four different problems that use different simulation domains --- a fluid simulation, two finite element problems, and a hydrodynamics simulation.

All code was compiled for and executed on a machine with an Intel i7-4790 CPU and an Nvidia Titan Black, GK110 Kepler architecture GPU. Reference code was compiled with \ic{gcc} 4.9 with optimizations (-O3) enabled; Reference CUDA code was compiled with \ic{nvcc} 6.5. Along with total run times and code sizes, we also provide overhead of JIT compilation, which occurs once when the code executes for the first time, and the total memory used for storing data including constants, key fields, and hidden fields over relations.
\change{
To measure memory allocation, we instrument Ebb directly, use a CUDA profiler for reference GPU code, and a malloc counting tool for reference CPU code.
}

\subsection{FluidsGL}

FluidsGL~\cite{fluidsgl} is a simulation of the Navier-Stokes equations for incompressible fluid flow written in CUDA, that ships with the NVIDIA CUDA 6.5 SDK. We chose it as an example because it is simple, short and uses a regular grid, allowing easy hand-tuning on the GPU.  For these reasons, it serves as a good test for the quality of code Ebb produces. Furthermore, the simulation is based on an approach to Stable Fluids which requires a fast Fourier transform~\cite{stam1999stable}. FFTs are commonly used, heavily optimized, and exhibit communication patterns not well expressed as data parallel Ebb kernels.  To get good performance, Ebb code needs to interoperate with an external library for calculating them.

The code calculates the velocity of a fluid as values on a uniform grid. Each iteration, the fluid is transformed from the spatial domain to the frequency domain where the diffusion and projection steps are computed, and then transformed back into the spatial domain.  Point location is used to perform a cell-to-cell lookup when advecting velocity, and again for advecting a set of particles used to visualize the flow. In the reference implementation, the CUFFT library is used for the Fourier transforms.

\change{
Ebb also uses the CUFFT library to transform data into and out of the frequency domain---which is modeled by a second grid relation.  Using the Ebb API, we request a direct view of the GPU resident field data managed by the runtime.  In this way, CUFFT can operate directly on the memory, avoiding any additional cost of marshaling the velocity field data.
}

\begin{figure}
\begin{center}
 \includegraphics[width=\columnwidth]{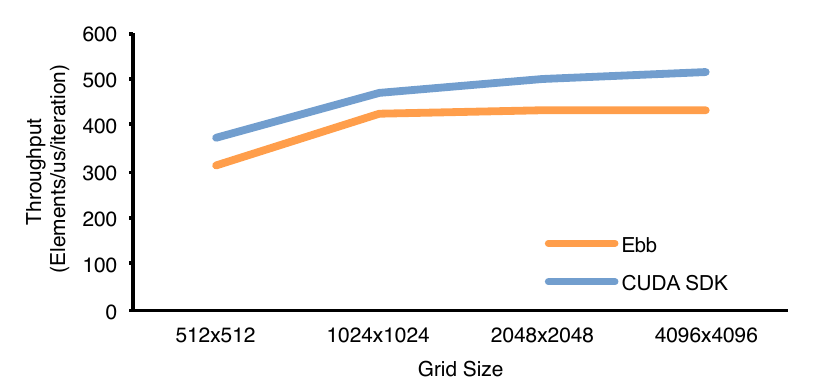}
 \caption{Our Ebb implementation of FluidsGL, an incompressible fluid flow simulation, compared against the performance of the implementation in the CUDA SDK.}
 \label{fig:FluidsGL}
\end{center}
\end{figure}

We compare the performance of our implementation to the original code (Figure~\ref{fig:FluidsGL}) across a range of grid sizes. Despite using a relational abstraction for managing the data, Ebb is able to produce code that runs no more than 19\% slower than the hand-optimized CUDA implementation. The total overhead for JIT compiling code is a constant 0.25 seconds, regardless of problem size.

Ebb's ability to interoperate with external libraries is important in this example. In both implementations, CUFFT accounts for around 30\% of the total compute time of each simulation iteration, a sizable but not dominant part of the runtime. Overall performance is a combination of both fast simulation code and fast FFT code. Providing interoperability allowed us to use the best implementations for each part of the application.

\change{
For problems sizes of $512^2$, $1024^2$, $2048^2$, and $4096^2$ cells, Ebb uses 30MB, 100MB, 370MB and 1300MB respectively, while the reference uses 30MB, 70MB, 250MB and 900MB.  Breaking this down into memory usage per-cell, Ebb uses approximately 80B-per-cell, while the reference uses approximately 53B-per-cell.  Taking a careful tally of the underlying problem, an optimal implementation can get away with a 32B-per-cell overhead. (16B for a double-buffered velocity field, 8B for frequency-domain storage and 8B for particle position---there is one particle for each cell)
Of the 48B of Ebb overhead, 16B is due to since-removed system inefficiencies, 24B is due to redundant representation of data within the user code, and a final 8B is unavoidable due to the programming model---specifically the need to maintain explicit key-fields from the cells to themselves (semi-Lagrangian lookup) and from the particles to the cells.
The reference code uses an excess $\sim$21B, at least 8B of which is attributable to superfluously storing particle velocity.  As evidenced by this breakdown, neither FluidsGL, nor the Ebb version were written with much attention to optimizing memory footprint.  While no greater degree of attention was paid to memory usage in the remaining comparisons, Ebb nonetheless consistently used half or less memory than the reference code.
}

\subsection{Vega}

Vega~\cite{vega} is a popular C/C++ physics library for simulating 3D elastically deformable solids.  It supports a variety of integrators, elasticity models, as well as both tetrahedral and hexahedral domains.  To make a comparison, we configured our Vega simulations to use implicit backward-Euler integration and the Saint Venant-Kirchoff elasticity model on a tetrahedral domain.  After slicing out the relevant code paths, we found that Vega used 2500 lines of code for its single-core implementation.  As such, we chose it as a demonstration of the performance Ebb can achieve on a larger program. Unlike the simpler FluidsGL, Vega's size makes it inherently more difficult and costly to optimize. While the library authors care enough about performance to have added a multi-core CPU implementation, no GPU implementation of it currently exists (to the best of our knowledge).

We wrote an implementation of the described Vega code path in Ebb.  In doing so, we took care not to exploit the opportunity to refactor code across VEGA's abstraction boundaries.  Details of the domain model topology are described in \S\ref{sec:domain_modeling}.  Position, velocity, force and displacement fields are stored on the vertices and material properties on the tetrahedra.  We store the stiffness matrix as a $3\times3$-\ic{double}-matrix-valued field on the edges, effectively recreating the custom sparse block matrix data structure coded in Vega out of relational primitives.

At each time step, Vega computes internal elastic forces, stiffness, and damping, constructing a linear system that is solved to compute vertex velocities and displacements, subject to external forces.  In Ebb, we solve this system using a Jacobi-preconditioned conjugate gradient solver (PCG) written entirely in Ebb kernels; as such, we can run this same PCG solver on the GPU.  When run on a single core, Vega uses a similar PCG solver; on multi-core we have Vega use the multi-threaded Pardiso solver from Intel MKL~\cite{pardiso} as well as the separate multi-threaded force model and integrator implementations. Solvers often have different performance characteristics than other simulation code. To ensure they run well in Ebb, we allow writers to optionally annotate kernels with underlying characteristics such as block size, which can increase performance by up to 2x for some solver kernels.

\begin{figure}
\begin{center}
 \includegraphics[width=3.25in]{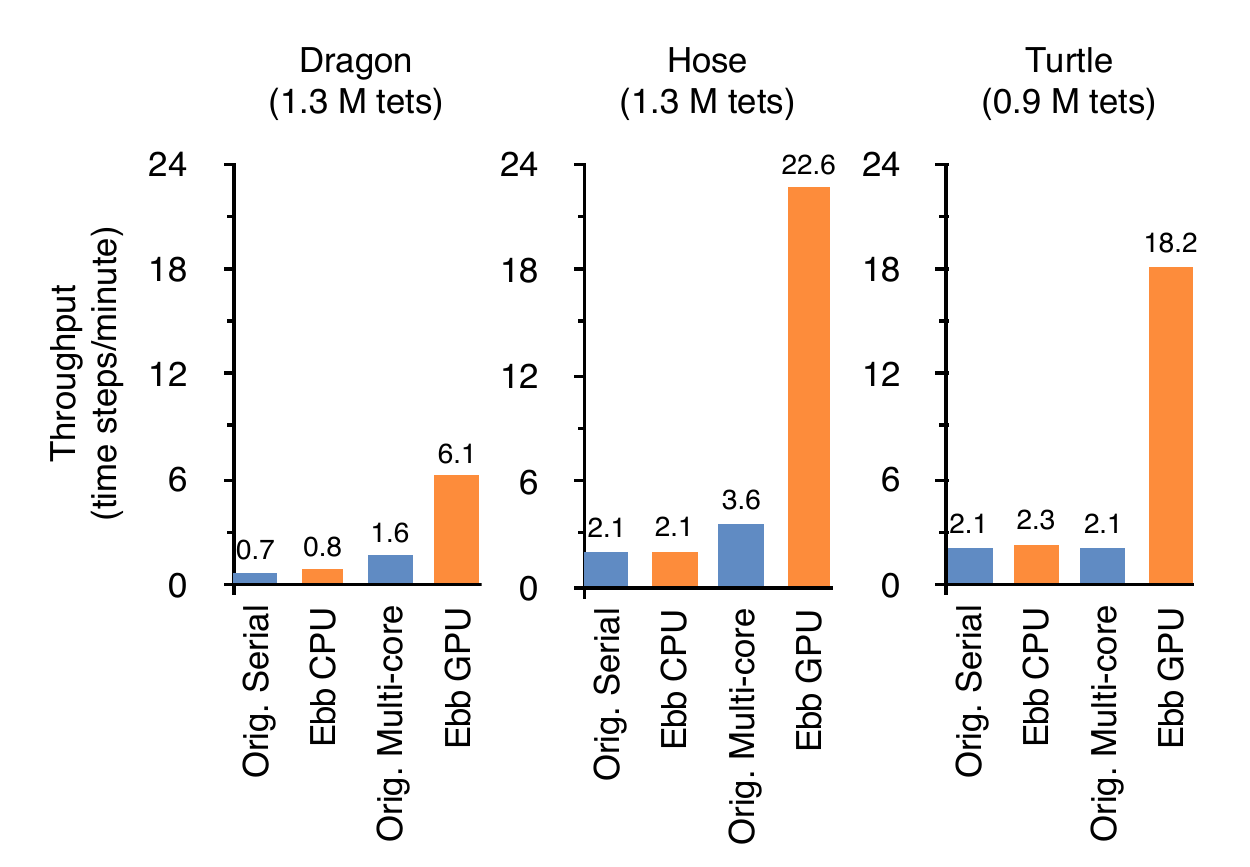}
 \caption{Ebb and reference average times per iteration, for 3 different meshes. Ebb GPU performs 9 times faster than reference serial, and 4 to 9 times faster than the best of serial and multi-threaded reference implementation over 8 cores.}
 \label{fig:vega-res}
\end{center}
\end{figure}

Figure~\ref{fig:vega-res} compares the performance of Vega in Ebb to the original Vega code for a few different meshes. The simulations in Ebb take 1.75 seconds to JIT compile once at the beginning, which is excluded from the figure. Note that our single core implementation of Vega matches the performance of the original code. We are also able to use Ebb's CUDA backend to generate a GPU implementation of Vega. Previously Vega could not run on GPUs. Our automatically-generated implementation runs 9 times faster than the serial implementation, and from 4 to 9 times faster compared to the best multi-threaded implementation on our CPU. For simulation library developers, this approach is much simpler and easier to maintain than translating reference code to GPU code by hand, which would require rebuilding data structures for GPU, rewriting the solver, and carefully considering the implementation of reductions to avoid race conditions.

Abstracting the domain via relations also simplified our implementation of Vega.
\change{
The original C/C++ code slice consists of over 2.5K lines of code for single-core computation, over 400 lines for loading tetrahedral meshes, and an additional 800 lines of code to implement multi-threading, not including the external solver.  In total VEGA takes over 3.7K codes to implement the exercised code paths. Our Ebb application is under 1K lines of simulation code, plus a 400 line tetrahedral mesh domain library, resulting in less than 1.4K lines total.
}
Ebb code is more concise due to automated memory management,
and the abstraction afforded by relational primitives
(encapsulated in macros).
The reference code, on the other hand, explicitly encodes the mesh as arrays augmented with specialized indexing structures for adjacencies. Furthermore, it requires the implementation of a separate sparse matrix class with fast indexing structures.
In our Ebb implementation, we were able to embed this sparse matrix structure as a matrix-valued field over edges, resulting in fewer lines of code. By avoiding redundant representations of the data, Ebb ends up using only 1.09GB, 1.06GB and 0.36GB for the dragon, hose and turtle meshes, compared to 2.35GB, 2.26GB and 1.46GB respectively for the reference code.

\subsection{FreeFem++}

FreeFem++ is a high level language designed for solving partial differential equations over meshes~\cite{freefem}. Problems in FreeFem++ are modeled using a variational formulation, which is closer to how physicists model partial differential equation problems. By comparing a FEM simulation in Ebb with one in FreeFem++, we evaluate the productivity vs performance tradeoffs of using a higher-level abstraction than the one Ebb offers.

We obtained a deformable simulation using a Neo-Hookean elasticity model, that was written in FreeFem++, with input from the FreeFem++ developers on the correct way to use their system. The simulation uses a conjugate gradient solver to perform implicit integration. We implemented the same elasticity model, with the same external conditions, numerically in Ebb.
Similarly to the Vega comparison, we store position, velocity, force and displacement on vertices, material properties on the tetrahedra, and stiffness matrix on the mesh edges. We reused our implicit integrator and conjugate gradient solver from the Saint Venant-Kirchoff example.  Counting both this integrator/solver and new code, our Neo-Hookean simulation written in Ebb requires about 800 lines of code. We also reused the tetrahedral mesh domain library from the Saint Venant-Kirchoff comparison, which was 450 lines of code. While typical FreeFem++ problems take tens of lines of code, the various tensor components introduced by the Neo-Hookean model made the FreeFem++ code about 700 lines long. 

We use FreeFem++ version 3.36 for evaluation. FreeFem++ takes about 41.3 seconds to run one time step on a tetrahedral mesh representing a sphere, with 2.4K tetrahedral elements. On a large bunny mesh with 78.7K tetrahedral elements, FreeFem++ takes about 28.3 minutes to run one time step. Ebb completes each time step on the sphere mesh in 0.006 seconds (6800 times faster) and on the bunny mesh in 0.26 seconds (6500 times faster), on a CPU. These meshes are too small to give any significant speedups on GPU---we get an additional speedup of 2 for the bunny mesh on GPU. The one-time overhead to JIT compile kernels is about 0.22 seconds when running on GPU and 0.4 to 0.6 seconds when running on CPU. Though Ebb can run this simulation on larger meshes, with an even larger speedup on GPU, we did not evaluate FreeFem++ with larger meshes due to the large running time.

Ebb performs better than FreeFem++ because the Ebb simulation computes elastic forces and stress using an algorithm that is specialized for the Neo-Hookean model. This is possible in Ebb, even with relatively concise code, because of the sufficiently low abstraction level that Ebb offers, while nonetheless relieving the user from the burden of memory management,
low-level data structure construction, and parallelization.

\subsection{Lulesh}

We also implemented a version of Lulesh (Livermore Unstructured Lagrangian Explicit Shock Hydrodynamics)~\cite{LULESH:spec}, which is a standard benchmark for evaluating the performance of simulation code across a variety of programming models. It has highly optimized implementations for a variety of languages and platforms~\cite{LULESH:versions}, allowing us to evaluate the quality of the code Ebb produces relative to highly-tuned implementations. Lulesh also uses a semi-structured hexahedral mesh, exercising another domain model.

\begin{figure}
\begin{center}
 \includegraphics[width=\columnwidth]{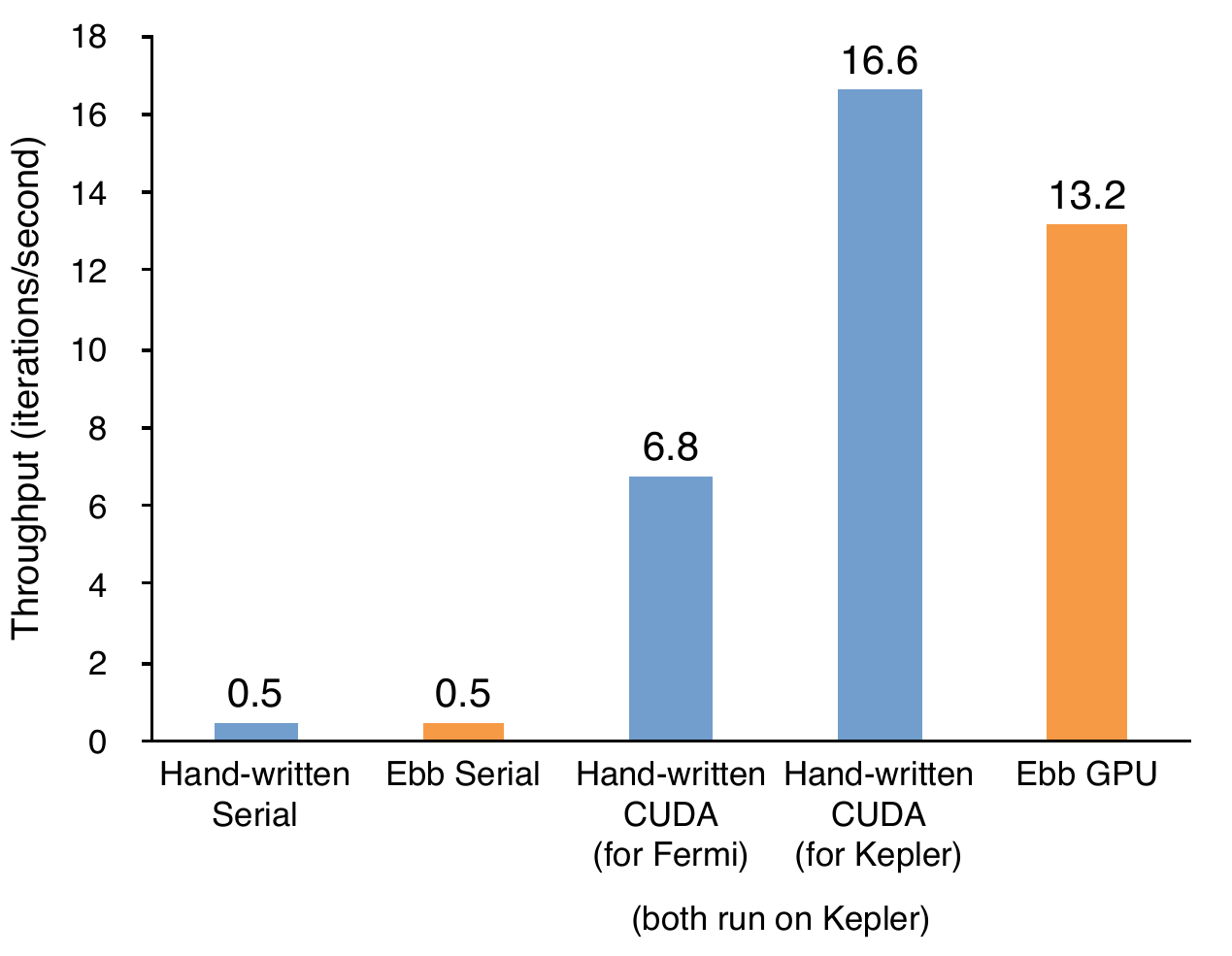}
 \caption{Ebb and reference GPU implementations of Lulesh, compared against the reference serial implementation. The simulation was run over a 150x150x150 hexahedral mesh for 2,527 iterations.}
 \label{fig:Lulesh}
\end{center}
\end{figure}

Lulesh models the propagation of a Sedov blast wave, using explicit integration. It stores thermodynamic variables, such as energy and pressure, mapped over hexahedral elements, and kinematic values, such as position and velocity, mapped over nodes. The iterative algorithm consists of a phase that advances node quantities, a phase that advances element quantities, and a phase that computes all values at the next time step.

Figure~\ref{fig:Lulesh} summarizes the results for Ebb compared with implementations of Lulesh that were hand-optimized for serial and GPU execution. The serial version of Ebb performs at the same speed as the hand-written reference. Our GPU implementation runs about 24 times faster than the serial code, and within 27\% of the performance of GPU code hand-tuned specifically for the Kepler architecture. Performance characteristics of GPUs can change over generations. The original CUDA version of Lulesh code was written for the previous Fermi GPU architecture and performs worse than the Ebb implementation. One advantage of writing simulations at a higher level is that changes in architecture can be handled by the compiler rather than by hand.  For instance, the addition of new atomics or changing best practices for reductions can be addressed within the Ebb compiler.

The time to JIT compile Lulesh code in Ebb is about 1.5 seconds. Ebb uses 1GB memory, compared to 2GB memory by reference code, for the $150^3$ hexahedral mesh.

We perform well compared to other domain-specific language implementations. On a $45^3$ sized mesh, the Liszt\cite{liszt} language can run Lulesh at 176 iterations per second, while Ebb can run the same simulation at 340 iterations per second. Liszt struggles with the Lulesh benchmark since its coloring-based approach to reductions does not work well for compute-heavy kernels~\cite{LULESH:perf}. Furthermore, Liszt uses an unstructured mesh as its only built-in domain, making it difficult to express code that assumes each element is a hexahedron. The unstructured mesh model also requires more memory to represent (Liszt could not fit a $150^3$ mesh into memory on our GPU).

Ebb makes implementing Lulesh easier compared to hand-written GPU models.  Ebb code (1.3K lines) is less than half the size of the GPU implementations (around 3.5K lines), and about the same size as a serial implementation (2K lines). This difference is largely because Ebb automatically handles synchronization, parallel reductions, data movement, and selection of block sizes for CUDA kernels. While the reference CUDA implementations explicitly include block reduction code needed to compute a minimum time step, Ebb generates that code automatically. Additionally, Ebb's built-in library for modeling hexahedral grids simplifies mesh construction, while automatic memory management simplifies the application code, resulting in a further reduction of lines of code, compared to the reference serial code.

\section{Discussion}

We have shown a three-layered simulation framework: the top layer where simulations are expressed as computational kernels over geometric domains, a middle, domain library layer in which the structure and implementation of geometric domains such as a regular 2D grid or arbitrary polyhedral meshes is defined, and a bottom, runtime layer that efficiently manages memory, kernel compilation, and execution on both CPU and GPU architectures. Using this separation, we can express a core set of simulation problems in graphics and automatically compile these problems to GPUs, obtaining performance comparable to hand-written implementations. Simulation programmers are able to focus on the physics and algorithms using familiar syntax. As needed, additional geometric domains can be implemented, extending the space of problems supported by Ebb without having to implement new languages or runtimes for each new geometric domain.

The programming model provided by Ebb provides many opportunities for additional optimizations. Currently relational data is stored column-oriented. Row-oriented storage may be better for relations that are sparsely accessed, but whose fields are always used together. Furthermore, we currently strictly map one row of a relation to one GPU thread. In applications such as Vega or Lulesh, kernels may perform large intra-element matrix-like operations which consume a large number of GPU registers, reducing peak compute throughput. In these cases, using multiple threads to cooperate on matrix operations for a single row may reduce register pressure and improve performance. The choice of how to lay out data or how to run the kernel can be made orthogonally to the specification of a kernel, similar to how other DSLs such as  Halide~\cite{halide} allow the separate expression of schedules. Since compilation is performed dynamically in the Ebb model, it is also possible to auto-tune decisions about layout and execution by recompiling and timing kernels as the program runs.

Currently our implementation runs either entirely on the GPU or entirely on the CPU. In some contexts, such as real-time game engines, balancing where simulation code is run can improve cooperation with other tasks such as rendering which are competing for the same compute resources. The Ebb model is sufficiently abstract to permit dynamically changing the location of computation as needed.

In addition to improving the performance of our current programming model, we plan to investigate ways to support additional aspects of simulation in graphics. Currently our model focuses primarily on static topologies. (which we think of as ``topological queries'') For simulations that include collisions or use particles, there is also the need for geometric queries. We have some limited support for these queries already by supporting point location in regular grids. But we plan to incorporate more generic queries such as finding nearby or intersecting elements. To do so, Ebb will have to handle the construction, maintenance and use of acceleration structures such as octrees and bounding volume hierarchies that are critical for accelerating these queries.  Furthermore, while Ebb allows the user to add new relations and fields as the program executes, it does not currently support operations to insert or delete rows from an existing relation. Such fine-grained updates are necessary to support simulation techniques that rely on changing numbers of particles or adaptive remeshing.
\change{ Abstracting these features so that they efficiently generalize across different parallel architectures remains a significant challenge. }

We view Ebb as a step toward a more unified high-level system for implementing a wide range of simulations in graphics including fluids (Eulerian, Lagrangian, particle and explicit surface tracking methods), rigid bodies and cloth (with collisions), as well as fracture, and merging and splitting of elastic and plastic bodies.
Ebb is only that first step towards answering the question ``How can we tractably build systems that combine the simulation of diverse phenomena together in a single scene?''
In the long term, we should strive to create a system for simulation that is as high-performance and easy-to-use as rendering languages are today.

\begin{acks}
Ebb is part of the Liszt project and PSAAP2 center at Stanford University, and funded by the U.S.A. Department of Energy.  Additional funding is provided by the DOE Exascale Co-Design Centers, ExaCT and ExMatEx. The Liszt project has also received support from
the Stanford Pervasive Parallelism Laboratory. We thank our collaborators, especially Ivan Bermejo-Moreno and Thomas D. Economon for their feedback as regular users of the language.  We would also like to thank David Levin for providing us with the FreeFEM Neo-Hookean code we compared against.  Thank you to Michael Mara for help with figures.
\end{acks}

\bibliographystyle{acmtog}
\bibliography{ebb}


\end{document}